\documentclass{article}
\usepackage{hiph-art}
\usepackage{graphicx,wrapfig}
\usepackage{amssymb}
\volnumber{25} \issuenumber{1} \edyear{2006}                             
\frompage{000} \topage{000}                                              
\recrevdate{1 August 2006}                                              
\def\rightmark{Annihilation Mechanisms of $e^+e^-$ and $\mu^+\mu^-$ Production}
\def\leftmark{D.~Anchishkin}

\title{Annihilation Mechanisms of $e^+e^-$ and $\mu^+\mu^-$ Production \\
in Relativistic Nuclear Collisions }
\authors{
{D.~Anchishkin
\index{D.~Anchishkin} 
\index{D.~Anchishkin} 
}\\[2.812mm]
{\normalsize
\hspace*{-8pt} Bogolyubov Institute for Theoretical Physics,\\
03143 Kiev, Ukraine \\[0.2ex]
}}

\abstract{
\hspace*{12pt}
  The $\pi^+\pi^-$ and $q\bar{q}$
annihilation mechanisms of dilepton production during relativistic
nuclear collisions are studied.
We focus on the modifications caused by the specific features of
in-medium pion and quark states rather than by medium
modification of the $\rho $-meson spectral density.
The main ingredient emerging in our approach is a form-factor of the
multi-pion (multi-quark) system.
Replacing the usual delta-function the form-factor plays the role of
distribution which, in some sense, "connects" the total 4-momenta of
the annihilating and outgoing particles.
The difference between the c.m.s. velocities attributed to
annihilating and outgoing particles is a particular consequence of
this replacement and results in the appearance of a new factor in
the formula for the lepton pair production rate.
We obtained that the
form-factor of the multi-pion (multi-quark) system causes broadening
of the rate which is most pronounced for small invariant masses, in
particular, we obtain a growth of the rate for the invariant masses
below two masses of the annihilating particles.  }


\makeindex
\begin{document}

\maketitle
\setcounter{page}{1}

\section{Introduction}
\label{intro}
The main objective in studying strongly interacting matter, which is
formed during relativistic heavy ion collisions, is identification of
the quark-gluon plasma.
Lepton pairs and photons created during relativistic nucleus-nucleus
collision do not interact with the highly excited nuclear matter, they
leave the reaction zone without further rescattering.
That is why, the dileptons ($e^+e^-$ and $\mu^+\mu^-$ pairs) observed
in high-energy heavy ion collisions carry undistorted information
on the dense early stage of the reaction as well as on its dynamics.
The enhancement with an
invariant mass of $200\div 800$ MeV observed by the CERES
collaboration \cite{1,2} in the production of dileptons has received
a considerable attention (for the review, see Ref. \cite{3}).
It was found that a large part of the observed enhancement is due to
the medium effects (see Refs. \cite{4,5} and references therein).
Meanwhile, pion annihilation is the main source of dileptons which
come from the hadron matter \cite{6,7} (see also \cite{CERES-2005}).
That is why, the proper analysis of the dilepton
spectra obtained experimentally gives important data which
probe the pion dynamics in the dense nuclear matter.
The purpose of the present paper is to look once more on the
$\pi^+\pi^-$ and $q\bar{q}$ annihilation mechanisms of dilepton production
from the hadron and quark-gluon plasmas by accounting the
medium-induced modifications of the dilepton spectrum.
In order to do this, we concentrate on the modifications which are
due rather to intramedium pion and quark states,
than on the discussion of a modification of the $\rho$-meson
spectral density.
In accordance with our suggestions, the main features of a
pion wave function follow from the fact that pions live a finite time
in the system where they can take part in the annihilation
reaction.
As a consequence, the off-shell effects give an appreciable
contribution to the features of the annihilation process
specifically in the region of low invariant masses.
Moreover, if pions are the entities of a local subsystem, then the
spatial structure of the pion states is far from a plane-wave one and
this also gives the essential contribution to the features of the
dilepton spectrum.



\section{ Annihilation of particles in finite space-time volume}
\label{techno}

To carry out the outlined program, we assume that the pion liquid
formed after the equilibration exists in a finite volume, and the
confinement of pions to this volume is a direct consequence of the
presence of the dense hadron environment which prevents the escape
of pions during some mean lifetime $\tau$.
The same can be assumed concerning a hot system of quarks which are
confined to a quark-gluon droplet.
So, we assume the system of
pions (quarks) produced in high-energy heavy-ion collisions is
effectively bounded in a finite volume.
The pion (quark) wave
functions $\phi_\lambda(x)$, where $\lambda$ is a quantum number,
satisfy the proper boundary conditions and belong to the complete
set of functions.
For instance, the stationary wave functions may be taken as the
solutions of the Klein-Gordon equation
$\left( \mathbf{\nabla} ^2 +k^2 \right)\phi_\lambda({\bf x})= 0$,
where $k^2=E^2-m^2$, which satisfy the Dirichlet boundary condition
on the surface $S$: $\phi_\lambda({\bf x})|_S=0$.
For the box boundary, we get
$\phi_\mathbf{k}({\bf x}) =
 \sqrt{8/V} \prod_{i=1}^3  \theta(L_i-x_i) \theta(x_i) \sin{(k_i x_i)}
,$
where $V=L_1L_2L_3$ is the box volume, $\lambda\equiv {\bf
k}=(k_1,k_2,k_3)$, and components of the quasi-momentum run through
the discrete set $k_i=\pi n_i/L_i $ with $n_i=1, 2, 3, \ldots$
For the spherical geometry, the normalized solutions are written as
$\phi_{klm}({\bf r}) =
   \theta(R-r)  \left( 2/r \right)^{1/2}
  J_{l+1/2} (kr)
  Y_{lm}(\vartheta,\alpha)/
  RJ_{l+3/2} (kR)$,
where $\lambda=(k,l,m)$. Next, the field operators
$\hat{\varphi}(x)$ corresponding to the pion field should be
expanded in terms of these eigenfunctions, i.e.
\begin{equation}
\hat{\varphi}(x)
=
\int\frac{d^3k}{(2\pi)^32\omega_\mathbf{k}} \left[
a(\mathbf{k})\phi_\mathbf{k}(x)+b^+(\mathbf{k})\phi^*_\mathbf{k}(x)
\right] \, , \label{1}
\end{equation}
where $a(\mathbf{k})$ and $b(\mathbf{k})$ are the annihilation
operators of positive and negative pions, respectively.
On the other hand, the states corresponding to confined particles
can be written in a common way as
$\phi_\mathbf{k}({\bf x}) =
  \sqrt{ \rho({\bf x})/V } \, \Phi_\mathbf{k}({\bf x})$,
where, for instance,
$\sqrt{ \rho({\bf x}) }=
\prod_{i=1}^3 \left[\theta(L_i-x_i)\theta(x_i)\right]$
for a box and $\sqrt{ \rho({\bf x}) }=\theta(R-r)$ for a sphere,
respectively.
The function $\rho({\bf x})$ represents the information about the
geometry of a reaction region or cuts out the volume where the pions
(quarks) can annihilate.
Hence, for the evaluation
of $S$-matrix elements wave functions $\phi_\mathbf{k} (x)$
should be taken as the pion $in$-states once annihilating pions
belong to finite system.
The amplitude of the pion-pion annihilation to a lepton pair in the
first non-vanishing approximation is calculated via the chain
$\pi^+\pi^-\rightarrow \rho\rightarrow\gamma^*\rightarrow \bar{l}l$,
where the $\rho$-meson appears as an intermediate state in
accordance with the vector meson dominance.
The matrix element of the reaction is $\langle \mathrm{out} | S |
\mathrm{in} \rangle
 =  - \int d^4x_1\, d^4x_2 \,
   \langle {\bf p}_+,{\bf p}_- \left|
   \,  T \left[  {\it H}_I^\pi(x_1) \, {\it H}_I^l(x_2)  \right] \,
   \right| {\bf k}_1,{\bf k}_2 \rangle   $,
where ${\it H}_I^\pi(x)=-e\, j_\mu^\pi(x) \, A^\mu(x)$ and ${\it
H}_I^l(x)=-e\, j_\mu^l(x) \, A^\mu(x)$.
It is remarkable that the pion density $\rho({\bf x})$ appears as a
factor of the pion current. Indeed,
\begin{equation}
j^\pi_\mu(x) =
  -i \hat{\varphi} (x)
  \stackrel{\leftrightarrow}{\partial_\mu} \hat{\varphi}^+(x)
=
  \frac{\rho({\bf x})}{V}
  \left[ -i \hat{\Phi} (x) \stackrel{\leftrightarrow}{\partial_\mu}
     \hat{\Phi}^+(x) \right]
\, , \label{1a}
\end{equation}
where the field operator $\hat{\Phi} (x)$ is defined in the same way
as that in (\ref{1}) with the functions
$\phi_\mathbf{k}({\bf x})$ replaced by $\Phi_\mathbf{k}({\bf x})$.
Because of this
factorization, after the integration over the vertex $x$ the density
$\rho({\bf x})$ automatically cuts out the volume, where the
$\pi^+ \pi^-$ annihilation reaction is running.
At the same time, this
means that the density $\rho({\bf x})$ determines the volume of
quantum coherence, i.e. just the particles from this spatial domain
are capable to annihilate one with another and make contribution to
the amplitude of the reaction.
To obtain the overall rate, it is
necessary then to sum up the rates from every coherent domain of the
fireball.

For the sake of simplicity one can assume that the pion states can be
approximately represented as
$\phi_\mathbf{k}(x)=\sqrt{\rho(x)/V}e^{-i k\cdot x}$
($\rho(x)$ is the 4-density of pions in the volume $V$
where pions are in a local thermodynamic equilibrium).
In essence, this approximation considers just one mode
of the wave function $\Phi_\mathbf{k}({\bf x})$
and reflects the qualitative features of the pion states in a
real hadron plasma.
In the frame of this approximation
a simple calculation immediately shows that the $S$-matrix element
is proportional to the Fourier-transformed pion density $\rho(x)$,
i.e. $\langle \mathrm{out} | S | \mathrm{in} \rangle \propto
 \rho(k_1+k_2-p_+-p_-)$,
where $k_1$ and $k_2$ are the $4$-quasi-momenta of the initial pion
states and $p_+$ and $p_-$ are the $4$-momenta of the outgoing
leptons.
This means that the form-factor of the pion source
$\rho(k)$ stands here in place of the delta function which appears
in the standard calculations, i.e. $(2\pi)^4\delta^4(K-P)\rightarrow
\rho(K-P)$, where $K=k_1+k_2$ and $P=p_++p_-$  are the total
(quasi-) momenta of pion and lepton pairs, respectively.
An immediate consequence of this is a breaking down of the
energy-momentum conservation in the $s$-channel of the reaction,
which means that the total momentum $K$ of the pion pair is no
longer equal exactly to the total momentum $P$ of the lepton pair.
The physical interpretation of this fact is rather obvious: the
effect of the hadron environment on the pion subsystem which
prevents the escape of pions from the fireball can be regarded
during the time span  $\tau$ as the influence of an external
nonstationary field.
The latter, as known, breaks down the
energy-momentum conservation. From now, the squared form-factor
$|\rho(K-P)|^2$ of the pion system plays the role of a distribution
which in some sense "connects" in s-channel the annihilating and
outgoing particles instead of $\delta$-function.
Indeed, the number
$N^{(\rho)}$  of produced lepton pairs from a finite pion system
related to an element of the dilepton momentum space, reads
\begin{equation}
\left< \frac{dN^{(\rho)}}{d^4P}\right>=\int d^4K |\rho(K-P)|^2
\left< \frac{dN}{d^4Kd^4P}\right>
\, ,
\label{2}
\end{equation}
where
\begin{eqnarray}
\left< \frac{dN}{d^4Kd^4P}\right>
=
\int \frac{d^3k_1}{(2\pi)^3 2E_1} \frac{d^3k_2}{(2\pi)^3 2E_2} \,
     \delta^4(k_1+k_2-K)\, f_\mathrm{BE}(E_1)\, f_\mathrm{BE}(E_2)
\\
\times \int \frac{d^3p_+}{(2\pi)^3 2E_+} \frac{d^3p_-}{(2\pi)^3 2E_-} \,
     \delta^4(p_+ + p_- - P)|A_0(k_1,k_2;p_+,p_-)|^2
\, .
\label{3}
\end{eqnarray}
Here, $E_i=\sqrt{m_\pi^2+\mathbf{k}_i^2}$, $i=1,2$ for pions and
$E_i=\sqrt{m_l^2+\mathbf{p}_i^2}$, $i=+,-$
for leptons, respectively.
To obtain Eq.(\ref{2}), we represent the amplitude of the reaction as
\begin{equation}
\langle \mathrm{out} | S^{(2)} | \mathrm{in} \rangle
=
\rho(k_1+k_2-p_+-p_-) \, A_0(k_1,k_2;p_+,p_-)
\, .
\label{4}
\end{equation}
Note that not only the form-factor $\rho(K-P)$ contains
information about the pion system.
The amplitude $A_0$ carries also new important features, which are
related to the violation of the energy-momentum conservation in the
$s$-channel.
The latter is a consequence of the medium effects through a partial
confinement of the pion states inside fireball what results in the
breaking of the translation invariance.
Indeed, the pion-pion c.m.s. moves with the velocity
$\mathbf{v}_{K}=\mathbf{K}/K_0$, whereas the lepton-pair c.m.s.
moves with the velocity  $\mathbf{v}_{P}=\mathbf{P}/P_0$.
Hence, these two center-of-mass systems are "disconnected"{} now, that is
why any quantity should be Lorentz-transformed when transferred from
one c.m.s. to another.
Reflection of this is the appearance of the correction factor
$\left[1+\frac{1}{3}\left(\frac{(P\cdot K)^2}{P^2K^2}-1\right)\right]$
in the formula for the dilepton production rate (for details see
\cite{anch-2002,anch-2004}).

By the broken brackets in Eq.(\ref{2}), we denote the
thermal averaging over the pion quasi-momentum space.
Actually, we assume a {\it local thermal equilibrium} in the
multi-hadron (-pion) system.
Hence, the Green's function,
$D^<(x_1,x_2)=\langle \hat{\Phi}^+(x_2) \hat{\Phi}(x_1) \rangle,$
which appears after thermal averaging, can be represented as
\begin{equation}
D^<(x_1,x_2)
=
\int \frac{d^4k}{(2\pi)^4} \, e^{-i k\cdot x} \,  A(k) \,
f_\mathrm{BE}(k,X)
\, ,
\label{4a}
\end{equation}
where $X=(x_1+x_2)/2,$ $x=x_1-x_2,$
$f_\mathrm{BE}(k,X)=\{\exp{[\beta(X)(k\cdot u(X)-\mu(X))]}-1\}^{-1} $
is the Bose-Einstein distribution function, which depends on space-time
variables, $X=(X^0,{\bf X}),$
$\beta$ is the inverse temperature, $u(X)$ is the hydrodynamical
velocity and $\mu$ is the chemical potential.
For the ideal gas (infinite life time of the system) the spectral
function $A(k^0,{\bf k})$ indicates that all states are on the
mass-shell: $A(k)=4\pi \delta(k^2-m_\pi^2) \, \theta(k^0).$
In the interacting system the spectral function reflects a collision
broadening of the states which includes as well a global decay of the
system if collisions in the system exist during finite time span.
For instance, the fireball, which is nothing else as a system of
strongly interacting particles, lives until its decay, i.e. starting
from the creation till the freeze-out, after which there are no
strong interactions between particles at all.
In our further consideration we will take into account just a global
decay of the multi-pion (multi-quark) system.
Assuming a proper model of the spectral function, $A(k),$ which
responsible in the present approach for finite life time of the
system, we incorporate it together with the spatial density
$\rho({\bf x})$ to the global system form-factor $\rho(x)$.

Concerning the physical meaning of Eq.~(\ref{2}), we note that one
can regard it as the averaging of the random quantity
$\left<\frac{dN}{d^4Kd^4P}\right>$  with the help of the
distribution function $|\rho(K-P)|^2$ centered around the value
$P$, which is fixed by experimental measurement.
In this sense, the hadron medium holding pions in a local
spatial region for some time, which is expressed as the local pion
distribution $\rho(x)$, plays the role of an environment randomizing
the pion source.
This randomization is a purely quantum one in contrast to the thermal
randomization of the multi-pion system which is already included to
the quantity $\left< \frac{dN}{d^4Kd^4P}\right>$.

\section{ Dilepton emission rates  }

In order to transform the distribution of the number of created
lepton pairs over the dilepton momentum space to the distribution
over invariant masses, one has to perform additional integration
using $\left< dN^{(\rho)}/d^4P \right>$ from (\ref{2}), i.e.
$\left< \frac{dN^{(\rho)}}{dM^2} \right> = \int\frac{d^3P}{2P_0}
\left< \frac{dN^{(\rho)}}{d^4P} \right>$, where
$P_0=\sqrt{M^2+\mathbf{P}^2}$.
This results in:
\begin{eqnarray}
&& \! \! \! \! \! \! \! \! \!
\left<\frac{dN^{(\rho)}}{dM^2}\right> =
 \frac{\alpha^2}{3(2\pi)^8} \left(1-\frac{4m^2_e}{M^2}\right)^{1/2}
 \left(1+\frac{2m^2_e}{M^2}\right) |F_\pi(M^2)|^2
 \int\frac{d^3P}{2P_0}\int d^4K \frac{K^2}{M^2}
\nonumber \\
&& \! \! \! \! \! \! \! \! \! \!
 \times
 \left|\rho(K-P)\right|^2 e^{-\beta K_0}
 \left(1-\frac{4m^2_\pi}{K^2}\right)^{3/2}\left[1+\frac13
 \left(\frac{(P\cdot K)^2}{M^2K^2}-1\right)\right]
\, ,
\label{5}
\end{eqnarray}
where  we take the Boltzmann distribution
$f_{\rm BE}(E)\approx \exp(-\beta E).$
Note, that during integration with respect to a 4-momentum $K$ one
should keep the invariant mass of a pion pair,
$M_\pi=\sqrt{K^2},$ not less than two pion masses.
On the other hand, possible finite values of the distribution
$\left<\frac{dN^{(\rho)}}{dM^2}\right>$ below the two-pion mass
threshold can occur just due to the presence of the pion system
form-factor $\rho(K-P)$.
The factor in the square brackets on the r.h.s. of (\ref{5}) is a
correction which is due to the Lorentz
transformation of the quantity $(\mathbf{k}_1-\mathbf{k}_2)^2$ from
the dilepton c.m.s. to the pion-pion c.m.s.
This factor gives a remarkable contribution to the dilepton spectrum
for invariant masses below the two-pion mass value.
Its influence is especially pronounced for $e^+e^-$ production as was
shown in \cite{anch-2002,anch-2004}.

In Eq.(\ref{5}) the $\rho$-meson form-factor, $F_\pi(M^2),$ is a
vacuum one.
Actually, there are two ways to take into account
effects of the hadron medium: first, one can account for distortion
of the pion states caused by dense environment; second, one can look
for $\rho$-meson polarization effects during its passing through the
hadron environment.
In the present paper we choose the first way of
accounting for the medium effects (see also \cite{knoll-2000}).
Just to elucidate as much as possible the consequences of the
contraction of the pion states in the hadron medium and to prevent a
double counting we take the  vacuum $\rho$-meson form-factor.

For particular evaluations we take as a model of the pion system the
Gaussian distribution of the particles in space and the Gaussian
decay of the system (this form-factor succeeded in HBT interferometry):

%
\begin{wrapfigure}{l}{0.32\textwidth}
{\includegraphics[width=0.3\textwidth,height=0.22\textheight,
                  angle=-90]{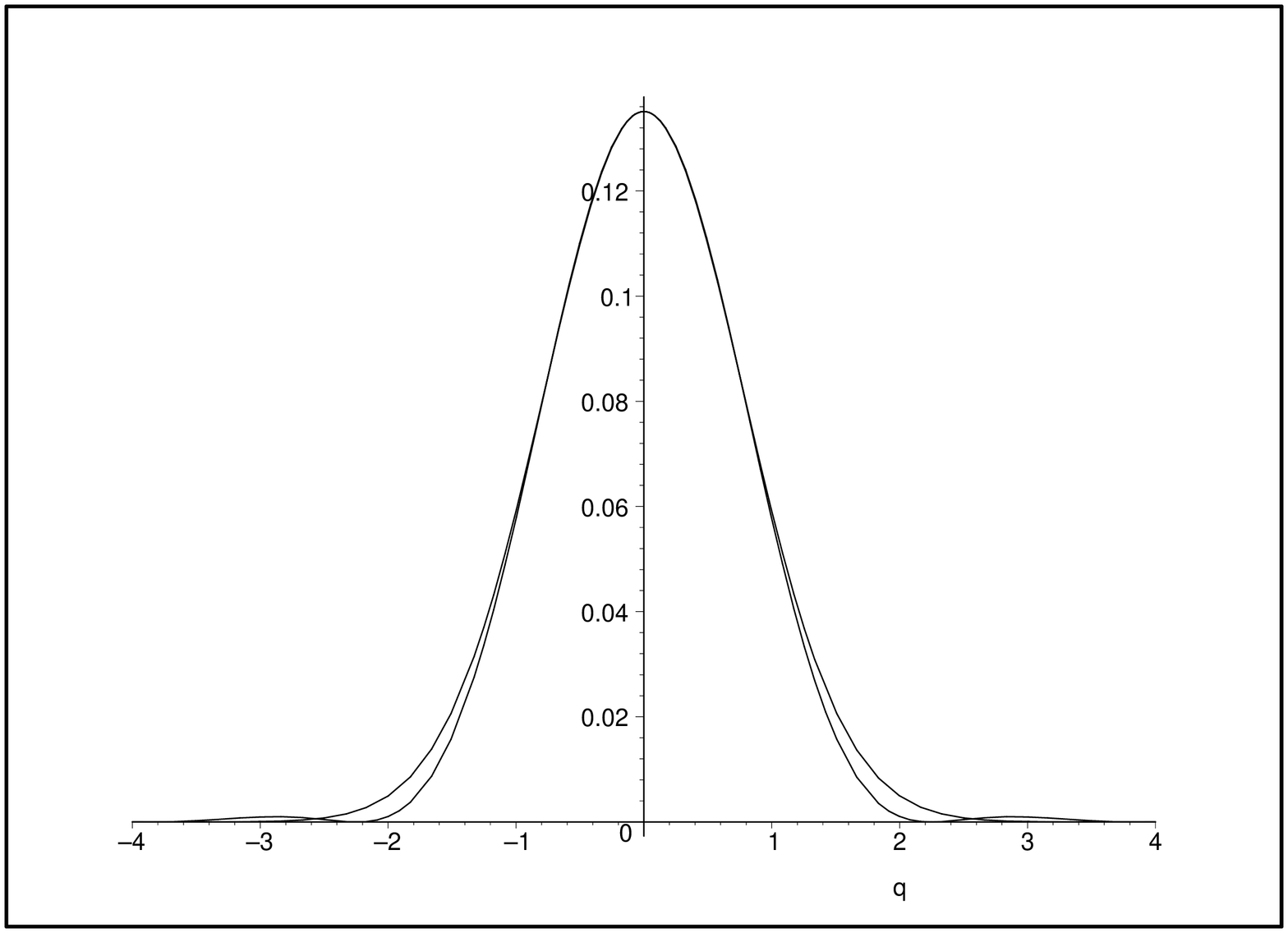}}

\vspace{30pt}
\noindent {\footnotesize Fig.~1. Comparison of the Gaussian and
$\theta-$function form-factors $|\rho(\mathbf{q})|^2/V$.}
\end{wrapfigure}

\noindent
\begin{equation}
\rho(x) =
\exp{\left( \frac{x^2-2(u\cdot x)^2}{2R^2} \right) }
\, ,
\label{6}
\end{equation}
where $u$ is the hydrodynamical (collective) velocity of the element
of the total system which is in a local thermodynamic equilibrium;
$R$ is the spatial radius of the element.
In the rest frame of the element the form-factor looks like:
$
\rho_0(t,\mathbf{r}) =
\exp{\left[ -(t^2+\mathbf{r}^2)/2R^2 \right] } .
$
To get a proper interpretation in terms of the mean life time of the
system element, $\tau ,$ one needs to make a scale transformation during
integration over the time variable:
$\int_{-\infty}^\infty dt \, \rho_0(t,\mathbf{r}) \, F(t,\mathbf{r})=
 \int_{-\infty}^\infty dt \, \exp{\left( -t^2/2\tau^2-\mathbf{r}^2/2R^2 \right) }
 F_0(t,\mathbf{r}),$
where $F_0(t,\mathbf{r})\\ =\frac{R}{\tau} F(\frac{R}{\tau} t,\mathbf{r}) .$

Meanwhile, it can be another choice of the pion source function.
Indeed, one can choose, for instance, a geometry with
sharp boundaries which are determined by the $\theta$-function.
To show that the final answer is not sensitive to the form of the
cutting function we compare two form-factors (normalized to
the unit volume) which correspond to the Gaussian distribution
$\rho(\mathbf{r})=\exp{ (-\mathbf{r}^2/2R^2) }$ and to the
$\theta$-function distribution
$\rho(\mathbf{r})=\theta(R-|\mathbf{r}|)$ (see Fig.~1).
Only a slight difference between these form-factors is seen and,
therefore, the choice of pion source distribution does not affect
much the dilepton production rate.

We evaluate the rate in the rapidity window,
$y_{\rm min} \le y \le y_{\rm max}$, which corresponds to CERES
experimental conditions \cite{1,2}
$$\frac{dR}{dM\, dy}=2\pi M\frac{1}{\triangle y}
\int_{y_\mathrm{min}}^{y_\mathrm{max}}\, dy
\int_{P_{\perp}\mathrm{min}}^\infty dP_\perp\, P_\perp
               \frac{dN}{d^4x\, d^4P}\, ,$$
where $\tanh{y}=P^3/P^0,\quad P_\perp^2=(P^1)^2+(P^2)^2$.
The results of evaluation of the production rates
$dR^{(\rho)}_{e^+e^-}/dMdy$ and $dR^{(\rho)}_{\mu^+\mu^-}/dMdy$ for
electron-positron and muon-muon pairs, respectively, in pion-pion
annihilation are depicted in Fig.~2.
Note, the calculations are carried out in the frame of the element of
the system where particles are in a local thermal equilibrium.
Different curves correspond to
the different "spatial sizes" $R$ and different "lifetimes" $\tau$
(for particular values of these parameters see Fig.~2) of a hot pion
system at the temperature $T=180$~MeV.


\begin{center}
\includegraphics[width=0.49\textwidth]{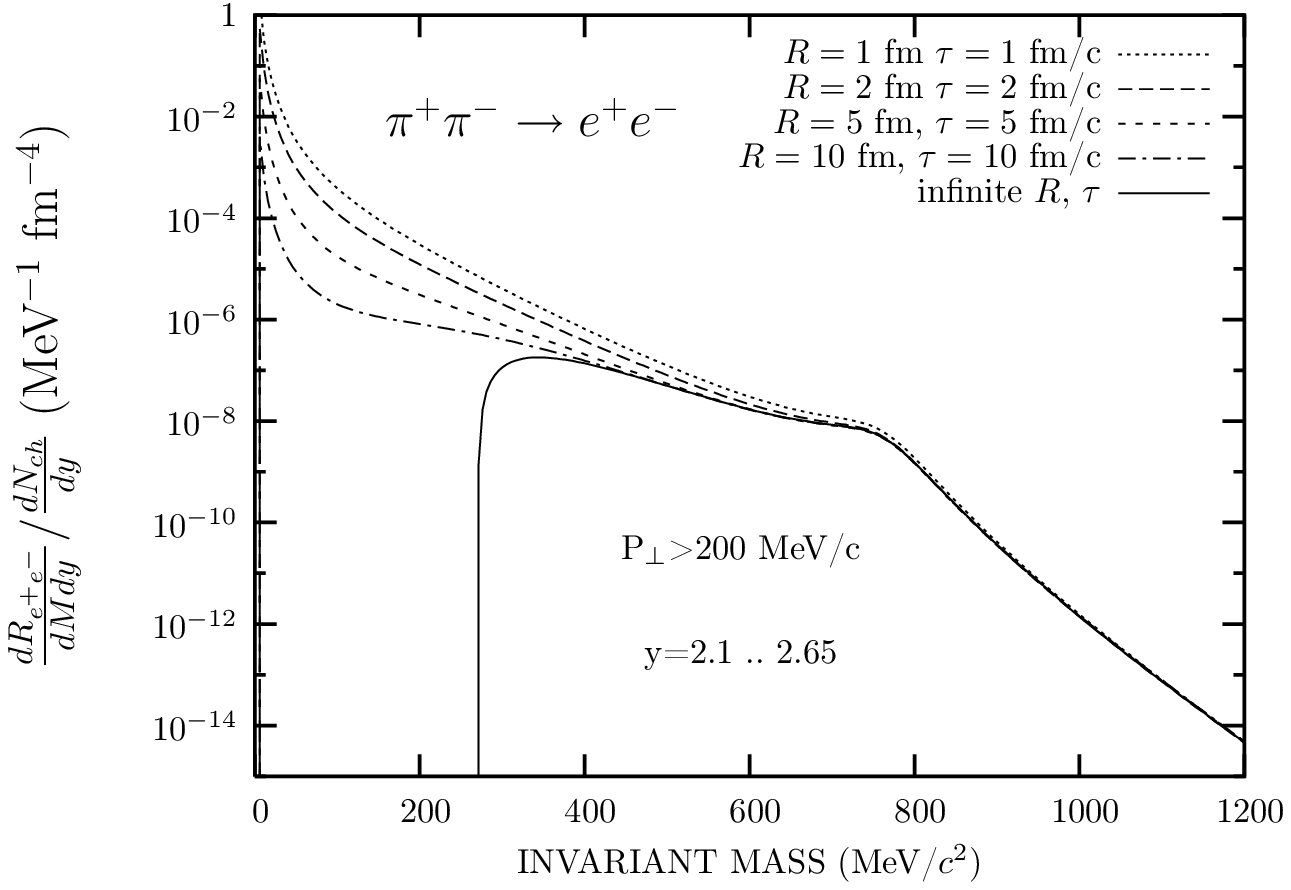}
\includegraphics[width=0.49\textwidth]{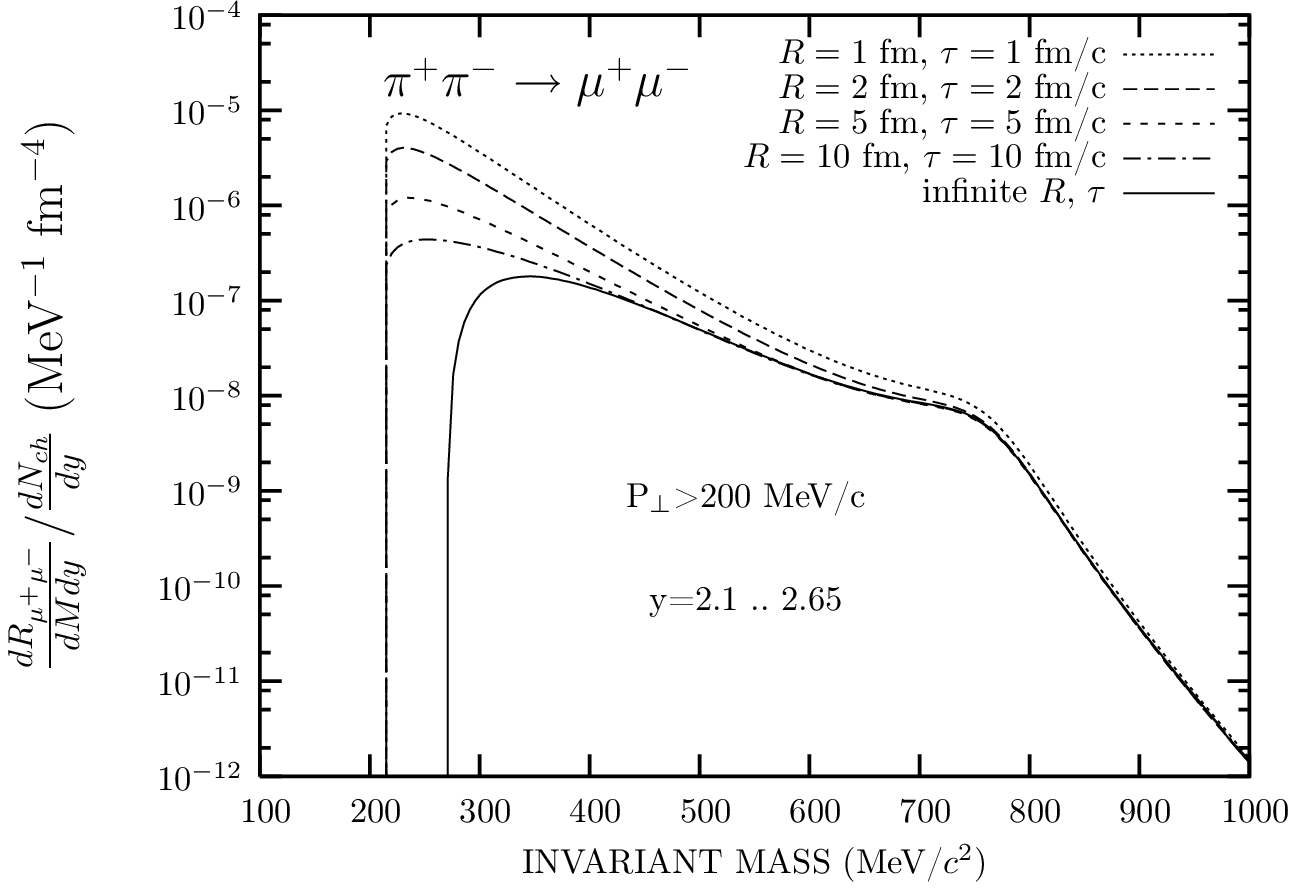}
\end{center}
\noindent {\footnotesize Fig. 2. Rates of electron-positron (left
panel) and muon-muon (right panel) productions in pion-pion
annihilation in a small finite system, $T=180$~MeV.}
\vspace*{10pt}%

For comparison, we present in Fig.~3 the results of evaluation of
the rate $dR^{(\rho)}_{e^+e^-}/dMdy$ of electron-positron pair
production in quark-antiquark annihilation
\begin{center}
\includegraphics[width=0.49\textwidth]{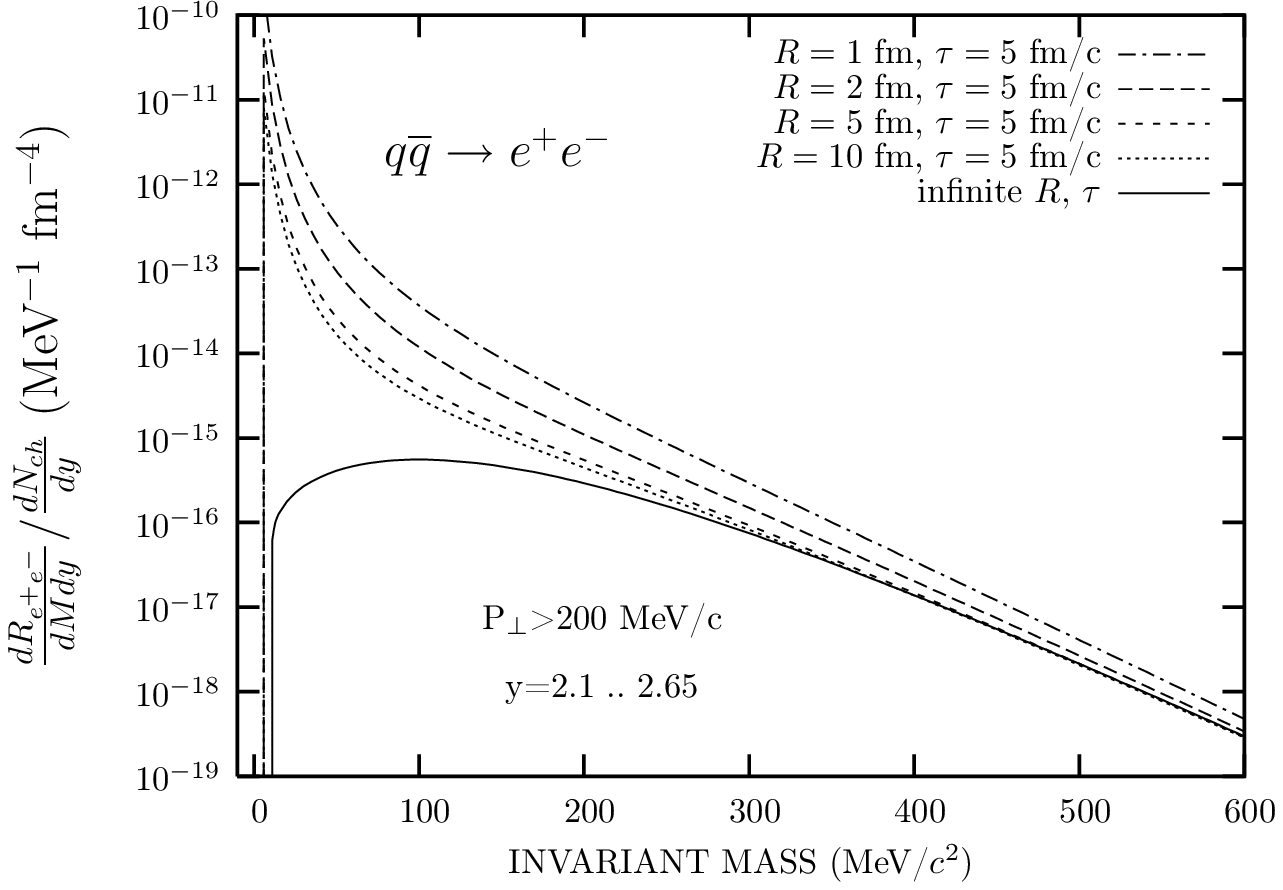}
\includegraphics[width=0.49\textwidth]{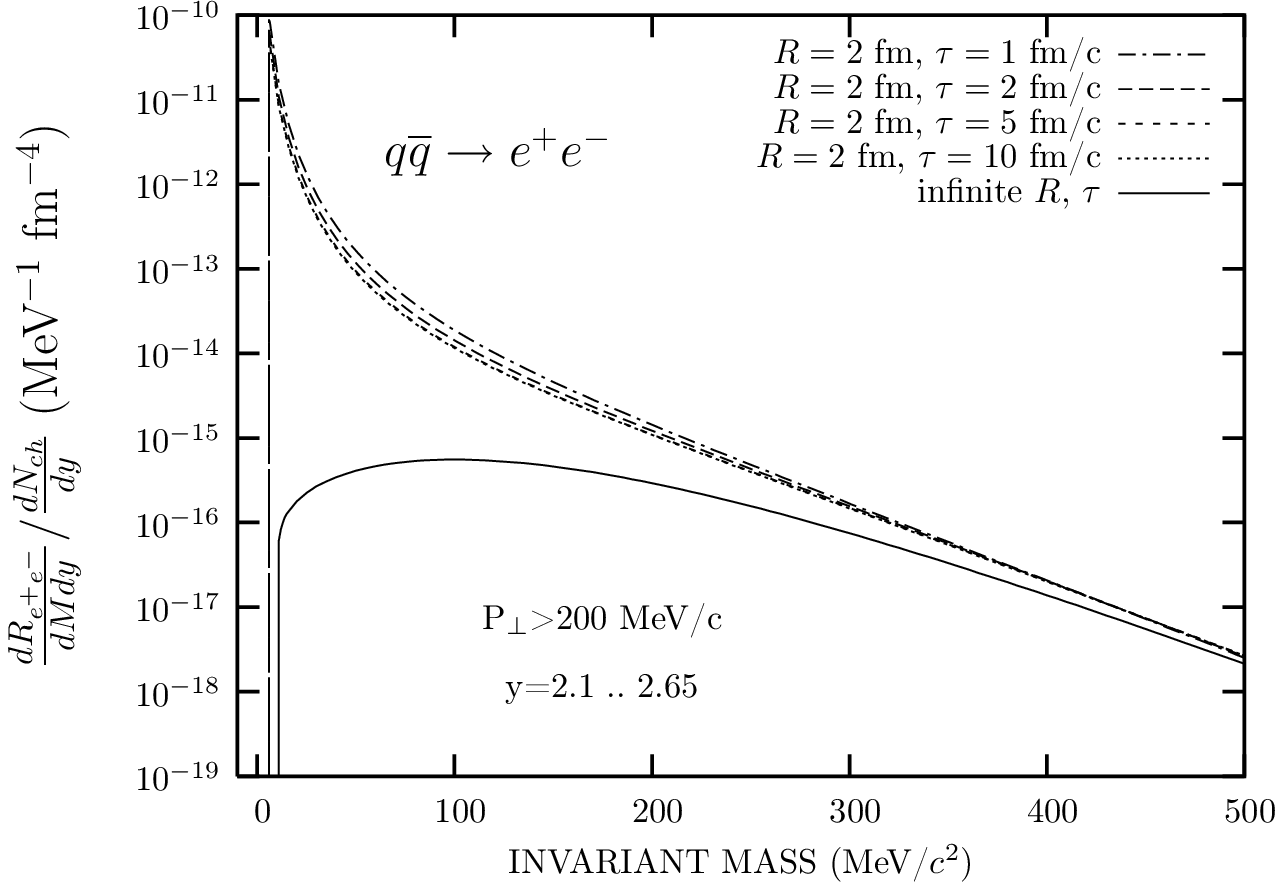}
\end{center}
\noindent
{\footnotesize Fig.~3. Electron-positron production rates
in quark-antiquark annihilation in a hot QGP drop. }
\vspace*{10pt}%

\noindent
in a hot QGP drop, $T=180$~MeV.
The evaluation was carried out in the frame of the quark drop under
the same assumptions as for pion-pion annihilation.
As in the previous case, an increase in the rate with decrease in the
invariant mass up to two electron masses is seen.
This real threshold is close to the total mass of annihilating quarks
$M=2m_q\thickapprox 10$~MeV/c$^2$.

\section{Conclusions}

We notice that the
production rate in a finite small system differs from the rate
in an infinite pion gas (solid curve) where pion $in$-states can be
taken as plane waves.
The deviation bigger when the parameters $R$ and $\tau$ are smaller.
Of course, this is a reflection of the
uncertainty principle which is realized by the presence of the
distribution $|\rho(K-P)|^2$ as the integrand factor in (\ref{2}).
Basically, the presence of the form-factor of the multi-pion system
will result in a broadening of the rate for small invariant masses
$M\leqslant 800$~MeV/c$^2$ which is wider at the smaller parameters
$R$ and $\tau$.
This seems natural because the quantum fluctuations
of the momentum are more pronounced in smaller systems.
We emphasize as well that the behavior of the curves in Fig.~2 which
correspond to a finite system has a similar to the CERES data
tendency \cite{1,2}.

The same behavior of the rate is seen in a hot quark drop (see Fig.~3):
for small parameters $R$ and $\tau$
in the region of small invariant masses $M\leqslant 500$~MeV/c$^2$, as
compared to the rate for infinite parameters $R=\infty$, $\tau=\infty$,
is due to a rise of quantum fluctuations which are evidently bigger
for a smaller size of the QGP drop.

Note that the enhancement of the
dilepton production rate for the low invariant mass region is much
more sensitive to the variation in the spatial size of a
many-particle (pion, quark) system than to that in the system
lifetime (see Fig.~3, right panel) \cite{anch-2005}.

\vspace{-2mm}

\section*{Acknowledgment}
The author would like to express his gratitude to V.~Khryapa,
R.~Naryshkin and V.~Ruuskanen for fruitful collaboration.
The work was supported by the Ukrainian Ministry of Education and
Science under contract M/101-2005.

\vfill\eject


\begin{thebibliography}{99}

\bibitem{1}
G. Agakichiev et al., {\it Phys. Rev. Lett.} {\bf 75} (1995) 1272.

\bibitem{2}
G. Agakichiev et al., {\it Phys. Lett.  {\bf B422} } (1998) 405.

\bibitem{3}
C. Gale and K.L. Haglin, {\tt hep-ph/0306098}.

\bibitem{4}
W. Cassing and E.L. Bratkovskaya, {\it Phys. Rept.} {\bf 308} , 65
(1999).

\bibitem{5}
R. Rapp and J. Wambach, {\it Adv. Nucl. Phys.} {\bf 25}, 1 (2000).

\bibitem{6}
V. Koch and C. Song, {\it Phys. Rev.}  {\bf C54} (1996) 1903.

\bibitem{7}
K. Haglin, {\it Phys. Rev.} {\bf C53} (1996) R2606.

\bibitem{CERES-2005}
G. Agakichiev et al.,
{\it Eur.Phys.J.} {\bf C41} (2005) 475;
{\tt nucl-ex/0506002}.

\bibitem{anch-2002}
D. Anchishkin, V. Khryapa, and V. Ruuskanen, {\tt hep-ph/0210346}.

\bibitem{anch-2004}
D. Anchishkin, V. Khryapa, R. Naryshkin, V. Ruuskanen,
{\it Ukrainian Journal of Physics} {\bf 49} (2004) 1039.

\bibitem{knoll-2000}
H. van Hees, J. Knoll,
{\it Nucl.Phys.} {\bf A683} (2000) 369; {\tt hep-ph/0007070}.

\bibitem{anch-2005}
D. Anchishkin, R. Naryshkin,
{\it Mod. Phys. Lett.} {\bf A20} (2005) 2047; {\tt nucl-th/0407042}.


\end{thebibliography}
\end{document}